**Mechanism of spin ordering in $Fe_3O_4$ nanoparticles by surface coating with organic acids**

Enrico Bianchetti and Cristiana Di Valentin*

Dipartimento di Scienza dei Materiali, Università di Milano Bicocca,

Via R. Cozzi 55, 20125 Milano, Italy

## Abstract

Saturation magnetization values close to the bulk have been reported for coated magnetite nanoparticles with organic acids. The mechanism of this effect is not yet understood. Here we show that a previously proposed rationalization in Nano Letters 12 (2021) 2499-2503 was based on electronic structure properties that are not consistent with several existing density functional theory studies. Our study is based on a wide set of Hubbard-corrected density functional tight binding (DTFB+U) and hybrid density functional theory (HSE06) calculations on $Fe_3O_4$ nanocubes of more than 400 atoms. We provide a new explanation for the spin ordering in coated nanoparticles, through the investigation of spin-flipping phenomena. In particular, we show that the spin-flip of d electrons at octahedral $Fe^{3+}$ sites, which is confirmed to be more favorable near the surface, especially where atomic reorganization can take place such as at corner sites, can be hampered by the presence of adsorbed organic acids because they do not only limit the surface reconstruction but also allow for additional ferromagnetic superexchange interaction between octahedral Fe sites as a consequence of the carboxylates bridging binding mode. The proof-of-concept of this mechanism is given by a simplified model of the Fe(III) tert-butoxide dimer.

* Corresponding author: cristiana.divalentin@unimib.it

## 1. Introduction

Magnetic nanoparticles (NPs) have attracted much attention for diagnostic and theragnostic medical applications, due to their small size, which is comparable to cell length scales and allows the interaction and interference with biological processes, however minimizing adverse effects and, thus, paving the way to novel approaches [1]. Among them, iron oxides nanoparticles have been playing a predominant role in nanomedicine, as contrast agents for magnetic resonance imaging (MRI), magnetic hyperthermia sources or drug delivery vehicles [2,3,4,5,6,7], due to their biocompatibility, low cost, high saturation magnetization and versatile surface chemistry for easy functionalization [8,9,10]. Magnetite nanoparticles are the first inorganic materials that have been clinically tested and approved for commercialization [11,12].

Nanoparticles of different and controlled sizes and shapes have been prepared through a variety of synthetic routes [9,13,14,15,16,17,18,19]; however, cubic nanoparticles enclosed by six (100) facets are the most commonly observed and their shape is compatible with the cubic unit cell of the inverse spinel crystal structure of $Fe_3O_4$ [20]. The O anions are organized in a face-centered cubic arrangement. The iron cations are either $Fe^{2+}$ or $Fe^{3+}$ in a 1:2 ratio. All $Fe^{2+}$ ions occupy octahedral sites in a random distribution with half of the $Fe^{3+}$ ions, whereas the other half of $Fe^{3+}$ ions occupy tetrahedral sites. The (001) surface is the most stable magnetite surface [21], as confirmed by calculations of the Wulff construction [22]. Therefore, under thermodynamic control, it is expected to be the most exposed in nanostructures [23].

The magnetic ordering in bulk magnetite derives from this structure organization of the Fe cations [24,25]. Since the Fe-Fe distances are too large for direct exchange, the superexchange through the O anions (Fe-O-Fe) is dominant and controls the magnetic interaction: being the $Fe_{oct}$-O-$Fe_{tet}$ angle close to 125°, an antiferromagnetic alignment is expected between the $Fe_{oct}$ and the $Fe_{tet}$ sublattices, which leads to a net ferrimagnetic order since the $Fe_{oct}$ ions are twice the $Fe_{tet}$ ions. The superexchange interaction in the case of $Fe_{oct}$-O-$Fe_{oct}$ is ferromagnetic because the angle is 90°. The magnetic moment per unit formulas is of 4.1 $\mu_B$ [26].

The magnetic properties of nanoparticles differ from those of bulk magnetite. Bulk materials are characterized by multidomains separated by domain walls, whereas nanoparticles present a single domain magnetic configuration. Progressively reducing the nanoparticle size increases the surface-to-bulk ratio. Due to spin canting at the surface layers, saturation magnetization is then largely reduced [27,28,29].

However, saturation magnetization values close to the bulk (96 emu/g) have been reported for coated magnetite nanoparticles with organic acids (84 emu/g vs 46 emu/g for naked NPs), with practically no canting of the magnetic moments at the surface [30,31,32,33]. Different adsorbed ligands have been systematically investigated [34,35]. Since the capping molecules are not magnetic, the origin of the enhanced magnetization is still an open question. An improved crystallinity or the restoring of a bulk coordination environment around the surface Fe ion are often invoked as the possible reasons. However, the origin of the NP-adsorbates interaction and mechanism how it is affecting the magnetic properties are not clear. There is certainly a general consensus on the fact that surface spins seem to better align with the overall magnetization direction of the NP when the surface is coated with ligands, but such effect is certainly not yet understood [30,31,32,34,35].

In a combined experimental and theoretical work, the authors propose an explanation based on a set of density functional theory (DFT) calculations comparing electronic structure of bulk magnetite, unreconstructed bare (100) surface and capped surface with organic acids [33]. However, the level of calculations and the results presented are not consistent with the existing literature. First, it was proven that the $Fe_3O_4$ (100) surface undergoes a large reconstruction that was not considered in this work [36]. Secondly, several density functional theory calculations in other studies, using DFT+U or hybrid functionals [37,38,39,40,41,42], clearly show a gap at Fermi level in the density of majority spin states for the unreconstructed bare (100) surface in total analogy with what observed for bulk [43,44,45], but in contrast with the metallic character observed by Salafranca et al. [33], on which fact they explain the reduced magnetization at magnetite surfaces.

Certainly, quantum mechanics is the only viable way to unveil the fundamental mechanisms governing the behavior of magnetic materials, but only if accurate models and methods are used. For this reason, in the present study we investigate real cubic nanoparticles (not just flat surfaces) and we go beyond standard DFT using a higher level theory method, i.e. hybrid functional HSE06, which in the past has proven to be reliable to catch the proper structural, electronic, and magnetic properties of magnetite [42,45,46].

Our results strongly support the hypothesis that the high magnetism registered for coated nanoparticles, close to the bulk value, should not simply be accounted for by a higher crystallinity of the samples or by a reduced disorder at the surface but, instead, it is also the result of a deeper involvement of the ligands in the mechanism of magnetization. The atomic level understanding achieved in this study indicates that the adsorbed carboxylate ligands (here resulting from the dissociation of acetic acid) become involved in a ferromagnetic superexchange interaction with pairs of $Fe_{oct}$ ions, which is the reason for the enhanced overall resulting magnetization.

This hypothesis is also corroborated by analogous observations in several carboxylato-bridged transition metal compounds, where superexchange coupling mediated by the carboxylate group was evidenced by experimental magnetic measurements [47,48,49]. This interesting analogy proves that the study of the exchange interaction between metal centers is a crossing point among several research fields, which are the study of existing magnetic materials, the design of novel magnetic materials (e.g. based on metal organic framework structures, MOF), the search for molecular magnets and the elucidation of the role of polymetallic sites in proteins.

## 2. Methods

To investigate surfactant adsorption modes in terms of simulated annealing and final geometry optimization, SCC-DFTB (short for self-consistent charge density-functional tight-binding), calculations were carried out using the software DFTB+ [50]. The SCC-DFTB is an approximated DFT-based method that derives from the second-order expansion of the Kohn-Sham total energy in DFT with respect to the electron density fluctuations. The SCC-DFTB total energy can be written as:

$$E_{tot} = \sum_{i}^{occ} \varepsilon_i + \frac{1}{2}\sum_{\alpha,\beta}^{N} \gamma_{\alpha\beta}\Delta q_{\alpha}\Delta q_{\beta} + E_{rep} \quad (1)$$

where the first term is the sum of the one-electron energies $\varepsilon_i$ coming from the diagonalization of the corresponding Kohn-Sham Hamiltonian matrix (where all three-center integrals are neglected), $\Delta q_\alpha$ and $\Delta q_\beta$ are the induced charges on the atoms α and β, respectively, and $\gamma_{\alpha\beta}$ is a Coulombic-like interaction potential. $E_{rep}$ is a short-range pairwise repulsive potential. The neglecting of three-center contributions from the Kohn-Sham Hamiltonian enables the use of integral tables. More details about the SCC-DFTB method can be found in Refs. [51,52]. DFTB will be used as a shorthand for SCC-DFTB.

For the Fe-Fe, Fe-H and Fe-C interactions, we used the “trans3d-0-1” set of parameters [53]. For the O-O, H-O, H-H, O-C, H-C and C-C interactions we used the “mio-1-1” set of parameters [54]. For the Fe-O interactions, we used the Slater-Koster files fitted by us previously [55], which can well reproduce the results for magnetite bulk and surfaces from HSE06 and PBE+U calculations. To properly deal with the strong correlation effects among Fe 3d electrons [56], DFTB+U with an effective U-J value of 3.5 eV was adopted according to our previous work on magnetite [42,45,46,55,57,58,59]. The convergence criterion of $10^{-4}$ a.u. for forces and the convergence threshold on the SCC procedure of $10^{-5}$ a.u. were used during geometry optimization together with conjugate gradient optimization algorithm. DFTB+U molecular dynamics (MD) simulations were performed within the canonical ensemble (NVT) using an Andersen thermostat that simulates a temperature annealing process up to 500 K. The temperature profile is shown in Figure S1 in the Supplementary material. The total simulation time is 30 ps with a time step of 1 fs. The convergence threshold on the SCC procedure was set to be $5\times10^{-3}$ a.u..

Once a good minimum was obtained at this lower and cheaper level theory, hybrid density functional theory calculations (HSE06) were carried out using the CRYSTAL17 package [60,61] to investigate electronic and magnetic properties of both naked and coated nanoparticles (NPs). For the validation, against experimental data, of the standard hybrid functional HSE06 as a robust theoretical approach to describe structural, electronic and magnetic properties of magnetite system, please refer to Ref. 45 and corresponding Supporting Information, where we also analyzed the effect of reducing the fraction of the exact exchange, in comparison with B3LYP calculations and PBE+U calculations with different U values. The Kohn−Sham orbitals are expanded in Gaussian-type orbitals: the all-electron basis sets are H|511G(p1), C|6311G(d11), O|8411G(d1) (for NP oxygen atoms), O|8411G(d11) (for surfactant oxygen atoms) and Fe|86411G(d41), according to the scheme previously used for $Fe_3O_4$ [42,45,46,55,57,58,59]. The convergence criterion of 0.023 eV/Å for forces was used during geometry optimization and the convergence criterion for total energy was set at $10^{-6}$ Hartree for all the calculations. The same computational set-up is used for the toy model Fe(III) metoxide dimer, a simplified system of Fe(III) tert-butoxide dimer [62].

The cubic NP model (429 atoms with edge length of 1.5 nm) used for this investigation (see Figure 1) has been obtained from a larger one (1466 atoms with edge length of 2.3 nm) recently proposed by Liu et al. [46] by simply reducing the total number of atoms. Both these models are enclosed by six (001) facets, as observed in experiments [15,17], and present surface reconstruction, which consists in the transfer of six-coordinated iron atoms nearby a corner from octahedral to tetrahedral sites. For more details regarding this reconstruction process, please refer to Ref. 46. The total magnetization is reduced according to the formula, which determines the total magnetic moment of a magnetite system, as proven in Ref. 46:

$$m_{tot} = 5 \times \left[N\left(Fe_{Oct}^{3+}\right) - N(Fe_{Tet}^{3+})\right] + 4 \times \left[N\left(Fe_{Oct}^{2+}\right) - N(Fe_{Tet}^{2+})\right] \quad (2)$$

where $Fe_{Oct}^{3+}$ and $Fe_{Oct}^{2+}$ are $Fe^{3+}$ and $Fe^{2+}$ ions at octahedral sites, $Fe_{Tet}^{3+}$ and $Fe_{Tet}^{2+}$ are $Fe^{3+}$ and $Fe^{2+}$ ions at tetrahedral sites and $N$ is the number of the corresponding ions. Similarly to what happens in bulk magnetite, for $Fe_{Oct}^{3+}$ and $Fe_{Tet}^{3+}$ the high-spin $3d^5$ configuration gives an atomic magnetic moment of +5 and -5 $\mu_B$, respectively; for $Fe_{Oct}^{2+}$ and $Fe_{Tet}^{2+}$ the high-spin $3d^6$ electron configuration gives +4 and -4 $\mu_B$, respectively. The total magnetic moment $m_{tot}$ was found to be 288 $\mu_B$ for the cubic NP under investigation.

To mimic spin disorder phenomena, we have forced some $Fe_{Oct}^{3+}$ to give an atomic magnetic moment of -5 $\mu_B$ instead of +5 $\mu_B$, lowering the total magnetic moment. We named the energy difference between spin-up and spin-down solutions as $\Delta E_{Spin-Flip}$. After the spin-flip we allowed for a full atomic relaxation of the NP and we named the energetic gain associated to the relaxation $\Delta E_{Relaxation}$. We called the sum of these two contributions $\Delta E_{SF+Rel}$:

$$\Delta E_{SF+Rel} = \Delta E_{Spin-Flip} + \Delta E_{Relaxation} \quad (3)$$

We note that the spin-flip of one $Fe_{Oct}^{3+}6c-(\mathbf{3})$ and all (four) equivalent $Fe_{Oct}^{3+}6c-(\mathbf{3})$ ions reduces the overall total magnetic moment of the NP from 288 to 278 and 248 $\mu_B$, respectively. It affects the electronic structure according to the spin polarized projected density of states (PDOS) shown in Figure S2 in the Supplementary material. The spin-flip operation has been carried out at different organic acid coverage. Oleic acid, usually used as surfactant during NPs synthesis, was substituted by an acid with a shorter alkyl chain, the acetic acid (AA), to limit the computational cost during all simulations.

In order to verify the agreement with the Eq. (2) at different AA coverage, we performed a series of HSE06 calculations, where we fully relaxed the NP atomic positions while varying the $m_{tot}$ (see Figure S3 and S4 in the Supplementary material). The minimum total energy is registered for $m_{tot}$ 248 (or 278) and 288 $\mu_B$ for the spin-flipped and for the non-spin-flipped system, respectively, in perfect agreement with the output by Eq. (2).

For the toy model Fe(III) metoxide dimer we have computed both the ferromagnetic (FM) and antiferromagnetic (AFM) configurations. In the first configuration, both high-spin $3d^5$ $Fe^{3+}$ ions give an atomic magnetic moment of +5 $\mu_B$ for an overall magnetic moment of +10 $\mu_B$. In the second one, the high-spin $3d^5$ $Fe^{3+}$ ions give an atomic magnetic moment of +5 and -5 $\mu_B$ for a null overall magnetic moment. The $\Delta E_{AFM-FM}$ for this toy model is a quantity that corresponds to the $\Delta E_{SF+Rel}$ for the magnetite nanoparticles and is defined as:

$$\Delta E_{AFM-FM} = E_{AFM} - E_{FM} \quad (4)$$

where $E_{AFM}$ and $E_{FM}$ are the total energies of the antiferromagnetic and ferromagnetic configurations, respectively.

The adsorption energy per acetic acid molecule ($E_{ads}$) was calculated as follows:

$$E_{ads} = \left(E_{total} - E_{NP} - N_{CH_3COOH} \times E_{CH_3COOH}\right) / N_{CH_3COOH} \quad (5)$$

where $E_{total}$ is the total energy of the whole system (NP and adsorbed acetic acid molecules), $E_{NP}$ is the energy of the $Fe_3O_4$ NP, $N_{CH_3COOH}$ is the number of acetic acid molecules adsorbed and $E_{CH_3COOH}$ is the energy of one isolated acetic molecule. This formula provides a value for the adsorption energy that is normalized by the total number of acetic acids.

The band center of mass (COM) was computed using the formula [63,64,65]:

$$band\ COM = \frac{\int_{-\infty}^{E_F} E\rho(E)dE}{\int_{-\infty}^{E_F} \rho(E)dE} \quad (6)$$

where $\mathrm{E}$ is the energy, $\mathrm{E_F}$ the Fermi energy (which is set to 0), and $\rho(\mathrm{E})$ the electronic density of states.

## 3. Results and Discussion

The cubic model used in this study (see Figure 1) is enclosed by six (001) facets, as observed in many experiments [15,17], it is made of 57.7 $Fe_3O_{4.4}$ units (173 Fe and 256 O, with a slight excess of O), and, in agreement with our previous simulated annealing molecular dynamics study of magnetite nanoparticles [46], it presents four reconstructed vertexes and is characterized by an outer-shell layer containing only $Fe^{3+}$ ions and a core where $Fe^{2+}/Fe^{3+}$ ions alternate. In the ground state of the nanoparticle, in perfect agreement with what happens in bulk magnetite, the unpaired d electrons at the $Fe_{oct}$ sites are in the up configuration, whereas at the $Fe_{tet}$ sites are in the down configuration. The total magnetic moment can be obtained from the general formula in Eq. (2) (see Methods), as derived in our previous work [46]. Analogously to bulk magnetite, for $Fe^{3+}_{Oct}$ and $Fe^{3+}_{Tet}$ the high-spin $3d^5$ configuration gives an atomic magnetic moment of +5 and -5 $\mu_B$, respectively; for $Fe^{2+}_{Oct}$ and $Fe^{2+}_{Tet}$ the high-spin $3d^6$ electron configuration gives +4 and -4 $\mu_B$, respectively. The total magnetic moment $m_{tot}$ for the cubic NP under investigation is 288 $\mu_B$.

Spin-disorder phenomena may reduce this optimal magnetic moment per NP and can be simulated by forcing the spin-flipping of the 3d electrons (see Figure 2) at some chosen iron centers in the NP [66]. In particular, we model the spin-flip at various octahedral Fe sites inverting the magnetic order of the $3d^5$ electrons from spin up to spin down (see Methods for more information). These electrons are no more spin-aligned with those of the other $Fe_{oct}$ sites in the NP, but they become align with the 3d electrons at the $Fe_{tet}$ sites. This specific spin-flip process, of all the d electrons of one $Fe^{3+}_{Oct}$ ion, is certainly expected to be unfavorable in bulk magnetite, and indeed we compute an energy cost of +0.42/+0.67 eV ($\Delta E_{Spin-Flip}$ values at the bottom of Table 1) when the electrons of one $Fe^{3+}_{Oct}$ ion are flipped in a $Fe_3O_4$ bulk cell of increasing size (2, 8 or 16 $Fe_3O_4$ units). The increasing energy cost associated to a lower density of the "spin-defect" suggests that the larger the number of $Fe_{oct}$ and $Fe_{tet}$ ions involved in the ferromagnetic and antiferromagnetic interactions, respectively, the larger the overall stabilization. We expect that a positive trend for increasing supercell sizes, which, however, are computationally too costly. The question is: what happens if we evaluate the same spin-flip on $Fe^{3+}_{Oct}$ sites at the NP surface? Considering that surface spin canting is often observe in experiments, one would expect that the energy cost for spin-flipping might drop.

**Table 1.** Selected $Fe_{Oct}^{3+}$ sites in the cubic NP of Figure 1 and 2 (site numbering in bold from Figure 2) and $Fe_{Oct}^{3+}$ sites in bulk magnetite considered for the spin-flip mechanism investigation. Relevant information of the surrounding coordination, stoichiometry, and energetics are reported.

| Fe label | Nearby Fe atoms ($Fe_{Oct}$; $Fe_{Tet}$) | $Fe_3O_{4+x}$ units | $\Delta E_{Spin\text{-}Flip}$ (eV) | $\Delta E_{Relaxation}$ (eV) | $\Delta E_{SF+Rel}$ (eV) |
|---|---|---|---|---|---|
| $Fe_{Oct}^{3+}4c-(\mathbf{1})$ | 3; 1 | 57.7 | +0.25 | -0.08 | +0.17 |
| $Fe_{Oct}^{3+}4c-(\mathbf{2})$ | 2; 3 | 57.7 | +0.62 | -0.18 | +0.44 |
| $Fe_{Oct}^{3+}6c-(\mathbf{3})$ | 6; 3 | 57.7 | +0.27 | -0.26 | +0.01 |
| $Fe_{Oct}^{3+}5c-(\mathbf{4})$ | 3; 4 | 57.7 | +0.42 | -0.09 | +0.34 |
| $Fe_{Oct}^{3+}5c-(\mathbf{5})$ | 4; 3 | 57.7 | +0.21 | -0.07 | +0.14 |
| $Fe_{Oct}^{3+}6c-(\mathbf{6})$ | 6; 6 | 57.7 | +0.41 | -0.07 | +0.33 |
| $Fe_{Oct}^{3+}6c-bulk$ | 6; 6 | 2 | +0.42 | - | +0.42 |
| $Fe_{Oct}^{3+}6c-bulk$ | 6; 6 | 8 | +0.51 | - | +0.51 |
| $Fe_{Oct}^{3+}6c-bulk$ | 6; 6 | 16 | +0.67 | - | +0.67 |

Indeed, we observe a reduction of the computed energy cost to spin-flip $3d^5$ electrons at some specific $Fe_{Oct}^{3+}$ ions in the surface layers, especially when we allow for a full atomic relaxation of the NP ($\Delta E_{SF+Rel} = \Delta E_{Spin-Flip} + \Delta E_{Relaxation}$). In particular, this is true for those $Fe_{oct}$ ions that are characterized by a larger deficiency of $Fe_{tet}$ ions in the next coordination sphere of the spin-flipped ion, such as $Fe_{Oct}^{3+}4c-(\mathbf{1})$, $Fe_{Oct}^{3+}6c-(\mathbf{3})$ and $Fe_{Oct}^{3+}5c-(\mathbf{5})$ sites (see Figure 2 and Table 1). The energy cost at $Fe_{Oct}^{3+}6c-(\mathbf{3})$ sites drops down to +0.01 eV, which means that the spin up and the spin down configurations for this $Fe_{Oct}^{3+}$ ions are almost isoenergetic, which indicates an easiness to spin-disorder. For $Fe_{Oct}^{3+}4c-(\mathbf{1})$ and $Fe_{Oct}^{3+}5c-(\mathbf{5})$ sites the energy drops to +0.17 and +0.14 eV, respectively. On the contrary, more bulk like sites or sites whose coordination spheres involve a larger number of $Fe_{tet}$ than of $Fe_{oct}$, the energy cost for spin-flip at $Fe_{Oct}^{3+}$ ions remains high, as much as +0.44, +0.34 and +0.33 eV for $Fe_{Oct}^{3+}4c-(\mathbf{2})$, $Fe_{Oct}^{3+}5c-(\mathbf{4})$ and $Fe_{Oct}^{3+}6c-(\mathbf{6})$ sites, respectively. We rationalize this with a stronger superexchange antiferromagnetic interaction with the surrounding $Fe_{tet}$. Moreover, when forcing all $Fe_{Oct}^{3+}6c-(\mathbf{3})$ ions in the NP (four) to spin-flip, we calculated an energy cost $\Delta E_{SF+Rel}$ of +0.04 eV, which means that the overall cost of each spin-flip is additive (+0.01 eV × 4), see Table 2.

**Table 2.** Selected $Fe^{3+}_{Oct}$ sites in the cubic NP of Figure 1 and 2 (site numbering in bold from Figure 2) considered for the spin-flip mechanism investigation at different acetic acid (AA) coverage.

| No. AA | NP coverage | No. spin-flipped $Fe^{3+}_{Oct}6c-(\mathbf{3})$ | $\Delta E_{Spin\text{-}Flip}$ (eV) | $\Delta E_{Relaxation}$ (eV) | $\Delta E_{SF+Rel}$ (eV) |
|---|---|---|---|---|---|
| 0 | Naked | 1 | +0.27 | -0.26 | +0.01 |
| 0 | Naked | 4 | +0.23 | -0.19 | +0.04 |
| 1 | Low | 1 | +0.08 | -0.13 | -0.05 |
| 4 | Low (at corners) | 4 | -0.04 | -0.15 | -0.19 |
| 3 | High (at one corner) | 1 | +0.16 | -0.02 | +0.14 |
| 48 | Full | 1 | +0.11 | -0.01 | +0.10 |
| 48 | Full | 4 | +0.45 | -0.04 | +0.40 |
| No. AA | NP coverage | No. spin-flipped $Fe^{3+}_{Oct}5c-(\mathbf{5})$ | $\Delta E_{Spin\text{-}Flip}$ (eV) | $\Delta E_{Relaxation}$ (eV) | $\Delta E_{SF+Rel}$ (eV) |
| 0 | Naked | 1 | +0.21 | -0.07 | +0.14 |
| 48 | High | 1 | +0.31 | -0.03 | +0.28 |

The next step in our study is to investigate the effect of organic acids adsorption on the surface to find an atomistic-level explanation of the reduced spin disorder observed in coated magnetite nanoparticles, with saturation magnetization values close to those of bulk samples. To this aim, we adsorbed acetic acid (AA) on the NP surface, starting with one isolated molecule and considering different adsorption modes [67], as shown in the ball-and-stick representations in Figure 3. We considered dissociated bidentate (a), undissociated monodentate and H-bonded (b), undissociated bidentate (c) acetic acid on two five-fold coordinated $Fe_{oct}$ sites on one (001) facet of the NP model and dissociated chelate (d) acetic acid at a four-fold coordinated $Fe_{oct}$ site on one edge of the NP model. The largest adsorption energy, according to our DFTB+U calculations, is obtained for the dissociated bidentate (a) mode (-3.11 eV), which is more than 1 eV larger than that for the other adsorption modes considered. In the cases where acetic acid molecules dissociate, the released protons are adsorbed on superficial O atoms that are not directly bound to a $Fe_{tet}$ ion, since these O sites have been previously determined to be more reactive as proton acceptor or basic sites [68,69,70]. Because the dissociated bidentate mode is energetically favoured, in the rest of this study we will only consider it as the binding mode of the coating acetic acid molecules.

We now more closely investigate what happens when one isolated acetic acid molecule is adsorbed at the corner of the NP, in particular on two $Fe_{oct}$ sites directly bound the O ion at the corner. Because there are four corners of this type in the NP, we also put acetic acid molecules on the other three corners. We consider both situations (1AA and 4AA) as a low coverage regimes, since the acetic acid molecules are quite far apart from the other. The adsorption energy is -2.98 eV in 1AA and -3.23 eV per acetic acid molecule in 4AA. We must notice that the corner site undergoes a reconstruction with the O site moving away from the corner, breaking one of its bonds with one of the three $Fe_{oct}$ ions and resulting to be a bridging O at one side of the corner (see details in Figure 4a and 4c).

Since the cost for spin-flipping the $Fe^{3+}_{Oct}6c-(\mathbf{3})$ ion was found to be very low (only +0.01 eV), we decided to choose this site to investigate the effect of acetic acid adsorption on the spin-flip process. The aim is to understand the mechanism how the coating molecules can actually affect the energetics of spin-flipping. Actually, in the presence of one isolated acetic acid molecule (1AA), the spin-flip process costs +0.08 eV, but allowing for atomic relaxation in the new spin configuration, there is an energy gain of -0.13 eV, which reverses the sign of the overall energy release leading to an energy gain ($\Delta E_{SF+Rel}$) of -0.05 eV. Considering the 4AA model, we observe that the overall effect is additive since the $\Delta E_{SF+Rel}$ is now -0.19 eV, which is almost exactly four times what computed for 1AA. These results suggest that spin disorder phenomena, which are already strongly favored at the surface of NPs, can be actually further promoted in the presence of low density of coating molecules at defects sites.

The next question is then: do we observe a coverage effect? To answer this, we investigated a fully decorated NP (full coverage regime), where all the under-coordinated $Fe_{oct}$ ions on the facets, edges and corners are saturated by adsorbing forty-eight bidentate dissociated acetic acid molecules (48AA, as shown in Figure 4b). The average adsorption energy per acetic acid molecule is -1.96 eV. It is noteworthy that at a full coverage, we observe a reduction of the atomic reconstruction only at corner sites, where the exposed O ion keeps its three bonds with the nearby $Fe_{oct}$ ions (see Figure 4c and compare with Figure 4d). Besides that, the atomic rearrangements are rather limited as we can infer from the comparison of the simulated EXAFS spectra for naked and fully decorated $Fe_3O_4$ NP in Figure S5 in the Supplementary material. We wish to note that we have also performed a simulated annealing calculation at the DFTB+U level of theory (up to 500 K; see Methods for more information) and then cooled down the system again to allow for atomic rearrangement that might have some activation barrier, however no such events were observed.

Then, we investigated the spin-flip processes for this fully decorated NP model. We considered two types of octahedral Fe sites, $Fe^{3+}_{Oct}6c-(\mathbf{3})$ and $Fe^{3+}_{Oct}5c-(\mathbf{5})$, because the first is near a corner site and is the one with a negligible energy cost for spin-flip (+0.01 eV per site) in the naked NP, whereas the second is on the flat facet, far from the corner, but still is characterized by a quite low energy cost for spin-flip (+0.14 eV) in the naked NP. As opposed to the low coverage (1AA) situation, in the presence of a full monolayer of acetic acid molecules, we observe that the spin-flip of both the octahedral Fe sites considered becomes unfavourable. In the case of $Fe^{3+}_{Oct}5c-(\mathbf{5})$ site, the energy change for the Spin-Flip+Relaxation process ($\Delta E_{SF+Rel}$) in the naked NP is +0.14 eV, whereas at full coverage (48AA) it becomes +0.28 eV. In the case of $Fe^{3+}_{Oct}6c-(\mathbf{3})$ site, the $\Delta E_{SF+Rel}$ in the naked NP is +0.01 eV, at low coverage (1AA) is -0.05 eV, whereas at full coverage (48AA) it becomes +0.10 eV. Considering this process at full coverage for all the four structurally equivalent $\mathrm{Fe}^{3+}_{\mathrm{Oct}}6\mathrm{c}-(\mathbf{3})$, we observe that the overall effect is additive since the $\Delta\mathrm{E}_{\mathrm{SF+Rel}}$ becomes +0.40 eV, which is four times what computed for one Fe ion (+0.10 eV). This result indicates that the spin-flip process involves a single Fe ion and is independent from other Fe ions, with no long-range effects.

We must provide an explanation why a full coverage adsorption of organic acids affects the spin-flip processes at $Fe^{3+}_{Oct}$ sites in the surface layers and why it makes them unfavorable, similarly to what happens in bulk-like sites. One simple reason, which is often invoked, is that the presence of a full coating layer during the synthesis enhances the crystallinity of the NPs. To give ground to this statement, we have simulated and analyzed the EXAFS spectra of the naked and fully coating NPs and compared them (see Figure S5 in the Supplementary material). From these spectra we cannot highlight any evidence of improved crystallinity for the cubic NP model we proposed. Indeed, these models can hardly be deformed from their bulk-like shape and the small deformations are present both in the naked and in the fully coated NPs. Moreover, these deformations in the Fe-O and Fe---Fe distances or Fe-O-Fe angles do not correlate with the observation of superexchange in the models (see Figure S6 and corresponding discussion in the Supplementary Material). Therefore, we must find another reason why the presence of high density of adsorbed acetic acid affects the magnetic properties of the NPs. We propose that the reason is related to an induced extra superexchange ferromagnetic effect among $Fe_{oct}$ sites in the surface layers, which counteracts the surface spin canting processes. Indeed, it is well recognized in the literature that carboxylate ligands bridging dinuclear transition metal ions mediate superexchange coupling [47,48,49]. Similarly, bridging carboxylate groups from the acetic acid molecules on the NP surface could create additional paths for the ferromagnetic superexchange between $Fe_{oct}$ ions. In order to understand the origin of this extra ferromagnetic superexchange we have analyzed the variation of the O *p* energy levels with respect to the Fe *d* ones since this difference ($\Delta_{pd}$) is inversely proportional to the magnetic exchange coupling constant (J), according to equation [71]:

$$J \;\alpha\; \frac{2t_{pd}^4}{\left(U_d+\Delta_{pd}\right)^2} \qquad (7)$$

where $t_{pd}$ is hopping integral between *p* and *d* orbitals, $U_d$ is the Coulomb repulsion between two electrons in a *d* orbital, and the charge transfer energy $\Delta_{pd}$ is the energy difference between $O^{2-}$ *p* states eigenvalues and $Fe^{3+}_{Oct}$ *d* states eigenvalues, respectively. In this work, the $\Delta_{pd}$ for the NP systems is estimated with the energy difference between the center of mass (COM) of the $O^{2-}$ *p*-band and of the $Fe^{3+}_{Oct}$ *d*-band, as detailed in the Methods section above. We obtain a $\Delta_{pd}$ of 2.18 eV for the naked NP and of 1.98 eV for the fully coated NP. According to the equation (7), a reduction of $\Delta_{pd}$ (by 0.2 eV) is expected to enhance the magnetic exchange coupling constant, which is in perfect agreement with the stronger ferromagnetic superexchange interaction correlated to the higher spin-flip energy cost observed in our calculations for the fully coated NP.

The effect of bridging carboxylate ligands on the magnetic properties of Fe atoms can be tested on a toy model, i.e. the Fe(III) metoxide dimer, which is a simplified system of the Fe(III) tert-butoxide dimer [62], shown in Figure 5a. For this dinuclear $Fe^{3+}$ complex we have computed both the ferromagnetic (FM) and antiferromagnetic (AFM) configurations (see Methods for more information). Then, we have evaluated how the presence of two additional carboxylate bridges (Figure 5b) or of four additional water/ethanol molecules (Figure 5c and 5d, respectively) affects the relative stability AFM-FM. The $\Delta E_{AFM-FM}$ ($\Delta E_{AFM-FM} = E_{AFM} - E_{FM}$) for this toy model is a quantity that corresponds to the $\Delta E_{SF+Rel}$ for the magnetite nanoparticles discussed throughout the paper. In line with our prediction, the presence of additional bridging carboxylate stabilizes the FM solution, whereas the water and the ethanol molecules favour the AFM one.

To close the cycle and confirm the different behaviour of ethanol with respect to what obtained for acetic acid on the NP, we also adsorbed six ethanol molecules on the magnetite nanocube (Figure S7), i.e. high coverage at one corner, in perfect analogy to what done for acetic acid in Figure 4d. Three of these ethanol molecules were found to spontaneously dissociate forming H-bonded dimers (similarly to what observed for water on $Fe_3O_4$(001) surface [42,57,72,73]) with dissociated protons adsorbing on the same superficial O sites as those involved in the acetic acid dissociation. In the case of ethanol, we have computed a small negative $\Delta \mathrm{E}_{\mathrm{SF+Rel}}$ value (-0.02 eV), which confirms that the spin-flip process is slightly favoured. This result is in contrast with what observed for acetic acid with a $\Delta \mathrm{E}_{\mathrm{SF+Rel}}$ value of +0.14 eV. Therefore, these additional calculations have proven that an induced extra superexchange ferromagnetic effect among $Fe_{oct}$ sites in the surface layers is only triggered by adsorption of bridging molecules that electronically connect different $Fe_{oct}$ sites, such as acetate, but not of monocoordinated molecules, such as ethanol.

## 4. Conclusions

To conclude, we have first shown that previously proposed explanations for the high magnetization of coated magnetite NPs with organic acids are not always robust or satisfactory. One rationalization [33] is based on different electronic structure properties of bare and functionalized magnetite surfaces from DFT calculations, which however are in contrast with all the other existing studies in the literature and with the results here reported for the naked NP (see Figure S8a in the Supplementary material). Another explanation, which is commonly mentioned in the experimental papers, attributes the enhanced magnetization (i) to a higher crystallinity of coated NPs and (ii) to the restored full coordination of surface Fe ions. However, on one side (i) we have shown, through the simulation of EXAFS spectra (see Figure S5 in the Supplementary material), that coating does not appreciably alter the crystalline properties of our cubic NP model. On the other side (ii) we confirm that full coordination plays an important role because it does not allow for surface reconstruction, which is found to be crucial in favoring spin-flipping.

Besides this aspect, from our results another fundamental factor emerges: surface functionalization with carboxylic acids creates chemical bridges between $Fe_{oct}$ sites for an additional ferromagnetic superexchange interaction, which is an effective mechanism enhancing the overall NP magnetization. We provide proof of this mechanism using a toy model of Fe(III) tert-butoxide dimer and reporting experimental evidences for other carboxylato-bridged transition metal compounds [47,48,49]. Moreover, in order to shed light on the origin of this extra ferromagnetic superexchange, we investigated its relation with the charge transfer energy, i.e. the energy difference between $O^{2-}$ p and $Fe^{3+}_{Oct}$ d states, that is expected to be of inverse proportionality [71]. Indeed, the charge transfer energy is lower for the fully coated NP than for the naked one, leading to a higher ferromagnetic exchange coupling constant, in perfect agreement with the higher spin-flip energy cost observed in our calculations.

This interesting analogy between bimetallic complexes and the iron oxide compounds, as we mentioned in the introduction, proves that our study is a crossing point among several research fields, such as magnetic materials, molecular magnets and polymetallic sites in proteins, just to cite a few. Our results provide a rational basis for the understanding of spin ordering in magnetic nanoparticles by coating with organic acids. The approach proposed in this work is widely applicable to other nanosystems and can be used to explore different metals, different ligands, different system size, structure and spatial organization. The resulting insight will guide the design and optimization of such nanosystems for a broad range of biomedical and technological applications.

**Authorship contribution statement**

**Enrico Bianchetti**: Conceptualization, Methodology, Formal analysis, Investigation, Data curation, Writing – Review & Editing, Visualization. **Cristiana Di Valentin**: Conceptualization, Methodology, Formal analysis, Resources, Writing – Original Draft, Writing – Review & Editing, Visualization, Supervision, Project administration, Funding acquisition.

**Acknowledgments**

The authors are grateful to Lorenzo Ferraro for his technical help and to Hongsheng Liu for useful discussions. The project has received funding from the European Research Council (ERC) under the European Union's HORIZON2020 research and innovation programme (ERC Grant Agreement No [647020]).

**Supplementary material available**

Annealing temperature profile of the DFTB-MD simulations. The relative total energy as a function of the total magnetic moment ($m_{tot}$) for the $Fe_3O_4$ nanoparticle at different acetic acid coverage. Simulated EXAFS spectra and DOS for the $Fe_3O_4$ nanoparticle at different acetic acid coverage. Detailed structural parameters for the corner of the NP: naked and in presence of additional acetic acid and ethanol molecules. Ball and stick representation of the $Fe_3O_4$ cubic NP at ethanol high coverage at one corner.

**Data availability statement**

The data that support the findings of this study are available from the corresponding author upon reasonable request.

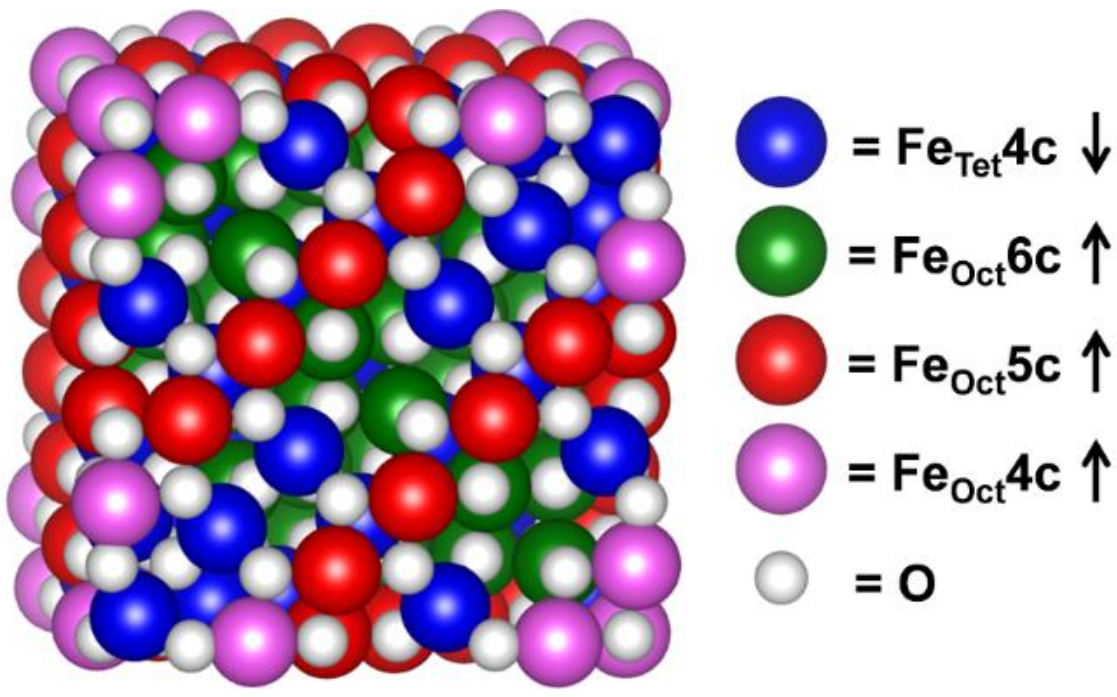

**Figure 1.** Schematic representation of the cubic nanoparticle model used in this study. The color coding of the atoms is given in the legend on the right.

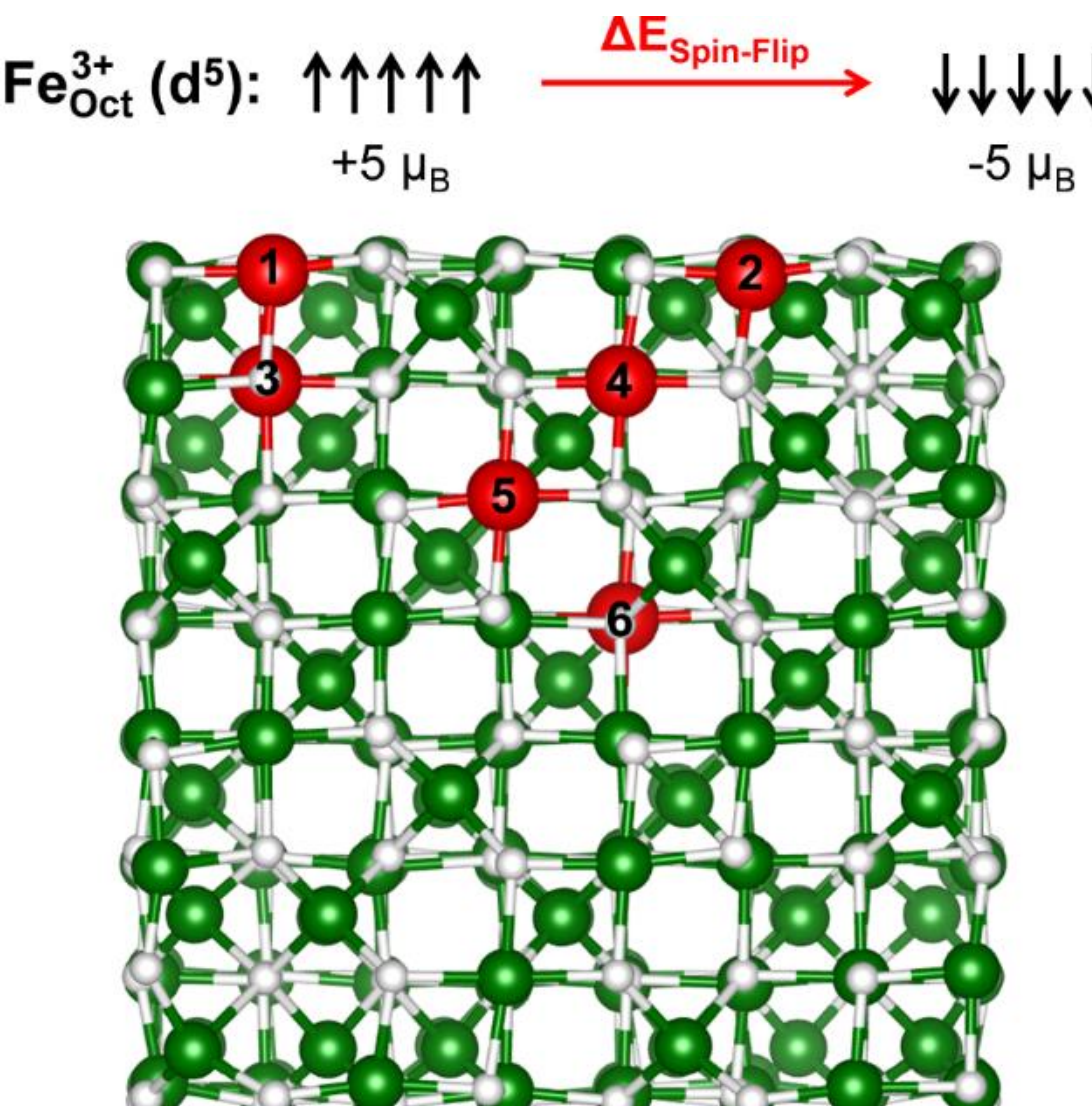

**Figure 2.** Schematic representation of the spin-flip mechanism (top) and global minimum structure of the magnetite cubic nanoparticle (bottom). The white, green, and red beads represent O, Fe, and $Fe_{Oct}$ on which the spin-flip cost is evaluated, respectively.

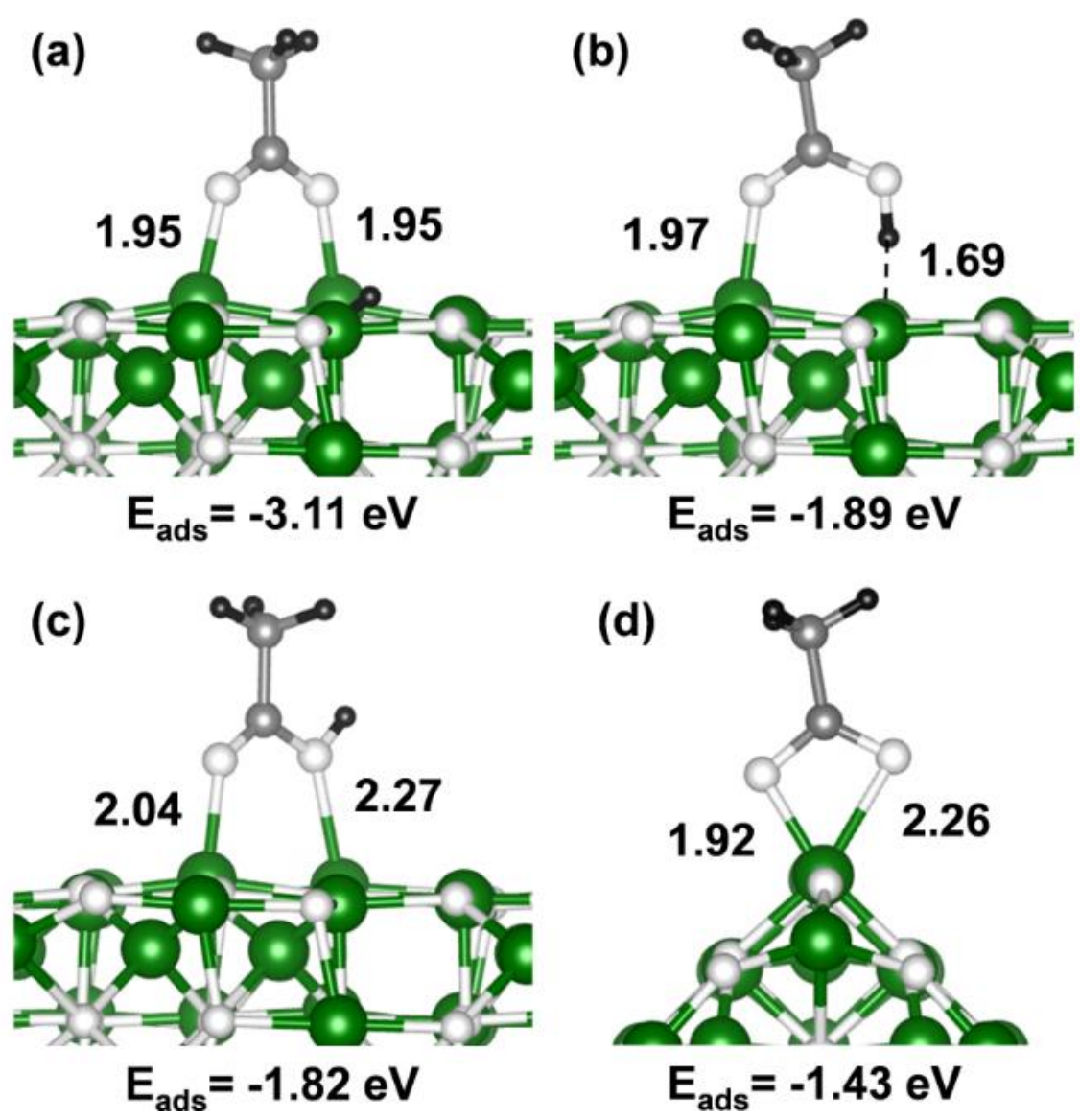

**Figure 3.** Ball and stick representation of the $Fe_3O_4$ cubic nanoparticle with an acetic acid molecule adsorbed on it in different configurations: (a) dissociated bidentate, (b) undissociated monodentate and H-bonded, (c) undissociated bidentate, and (d) dissociated chelate. The black, grey, white, and green beads represent H, C, O, and Fe, respectively. H-bonds are indicated by dashed black lines. The Fe-$O_{AA}$ bond length (in Å) and the adsorption energy (in eV) calculated at DFTB+U level are given for each configuration.

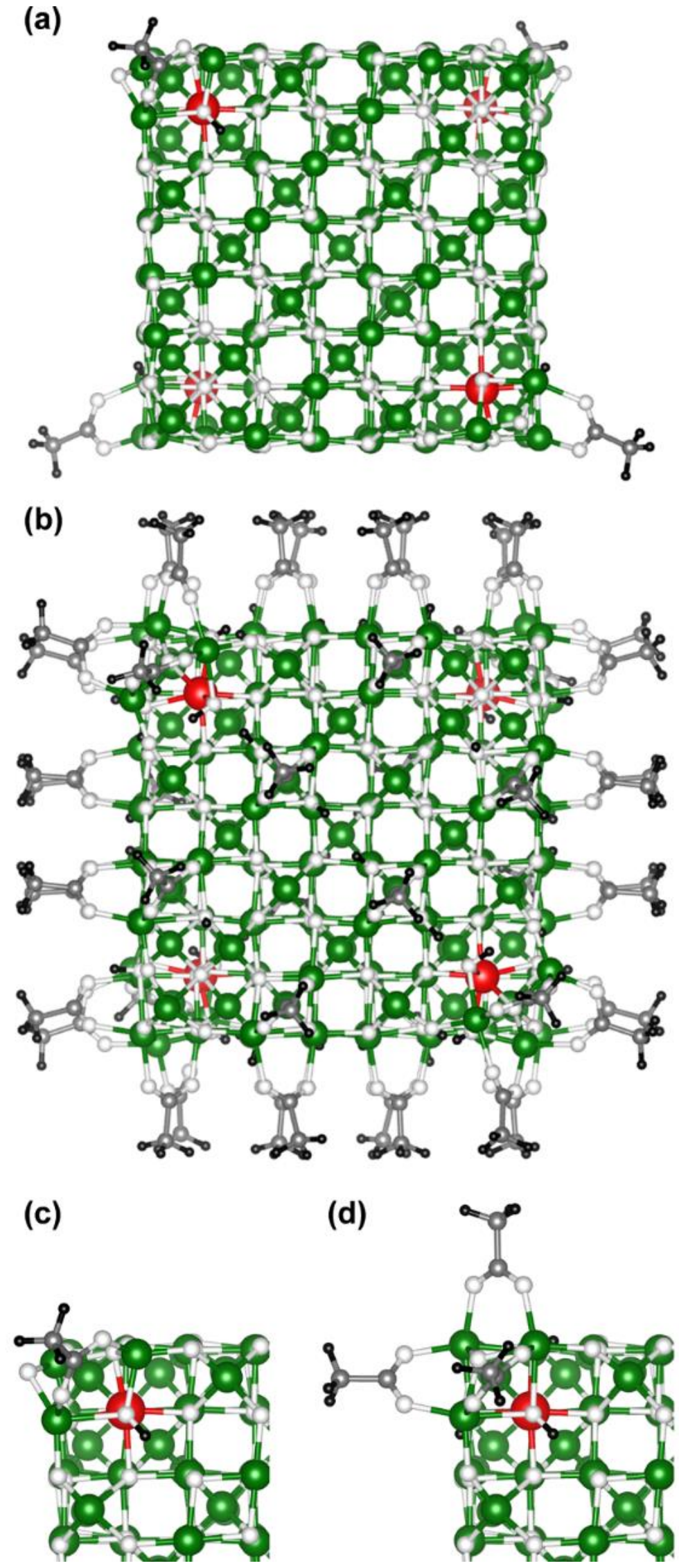

**Figure 4.** Ball and stick representation of the $Fe_3O_4$ cubic nanoparticle at different acetic acid (AA) coverage: (a) low coverage at corners (4AA), (b) full coverage (48AA), (c) low coverage (1AA), and (d) high coverage at one corner (3AA). The black, gray, white, and green beads represent H, C, O, and Fe, respectively. The red beads represent the $Fe_{Oct}$ on which the spin-flip cost is evaluated.

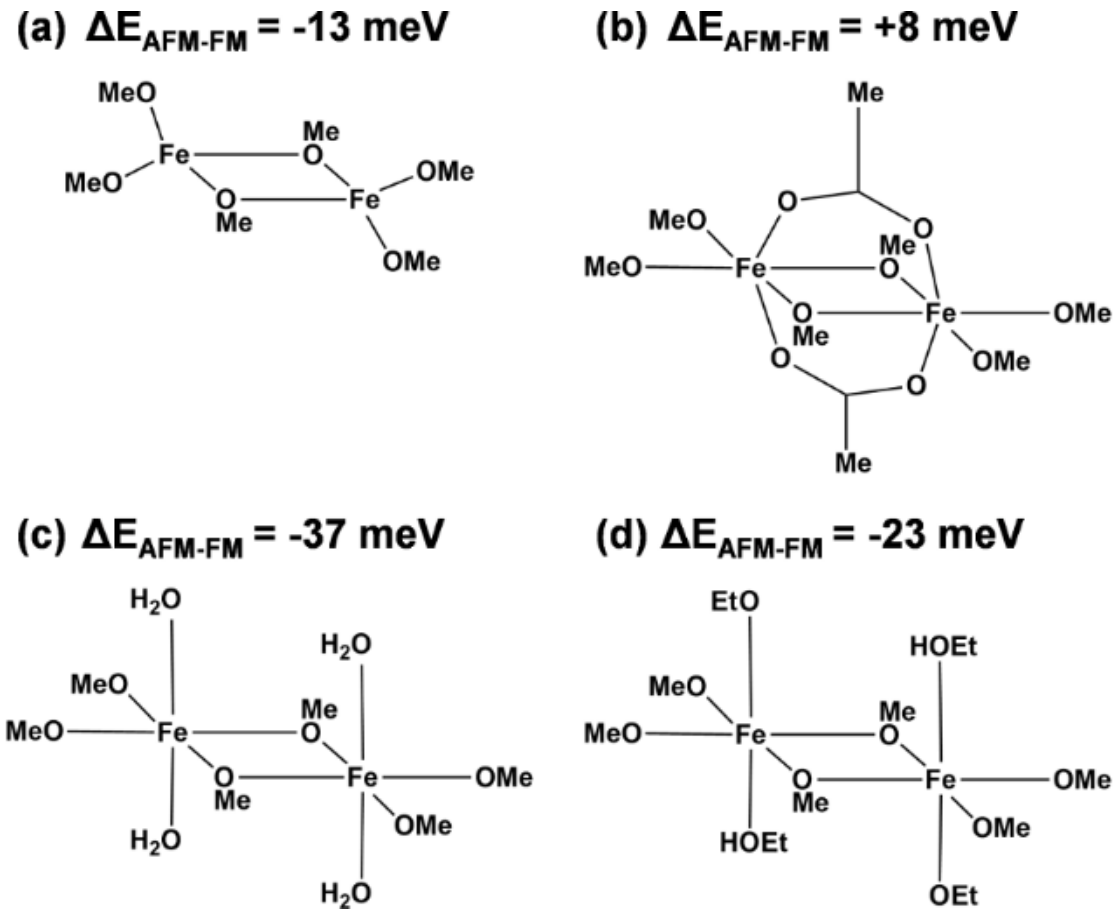

**Figure 5.** Schematic representation of the Fe(III) metoxide dimer (a) in vacuum, (b) with two adsorbed acetic acid molecules, (c) with four adsorbed water molecules, and (d) with four adsorbed ethanol molecules. The $\Delta E_{AFM-FM}$ calculated at the HSE06 level is given for each model.

**Supplementary material**

# Mechanism of spin ordering in $Fe_3O_4$ nanoparticles by surface coating with organic acids

Enrico Bianchetti and Cristiana Di Valentin

Dipartimento di Scienza dei Materiali, Università di Milano Bicocca,

Via R. Cozzi 55, 20125 Milano, Italy

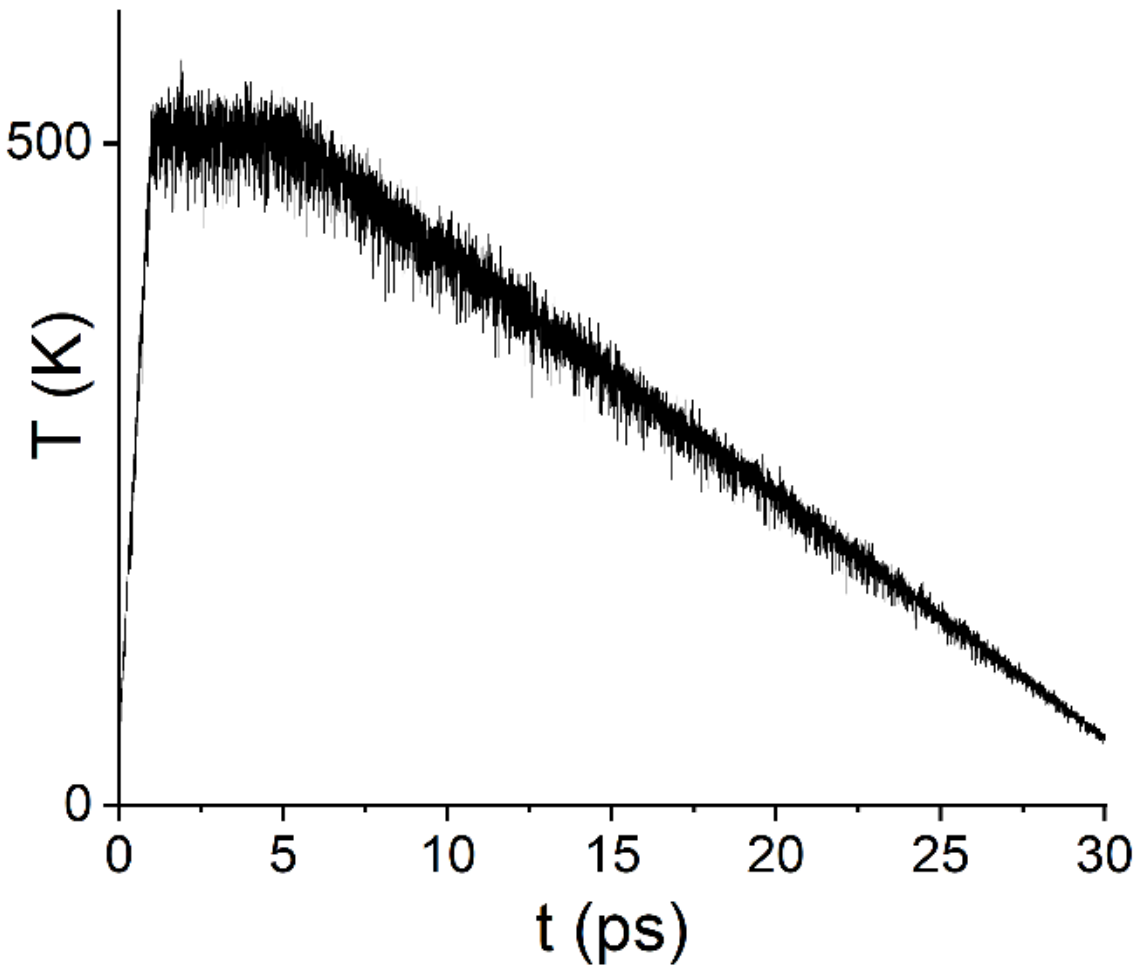

**Figure S1.** Temperature profile of the DFTB+U simulated annealing process for the NP at full coverage (48 AA).

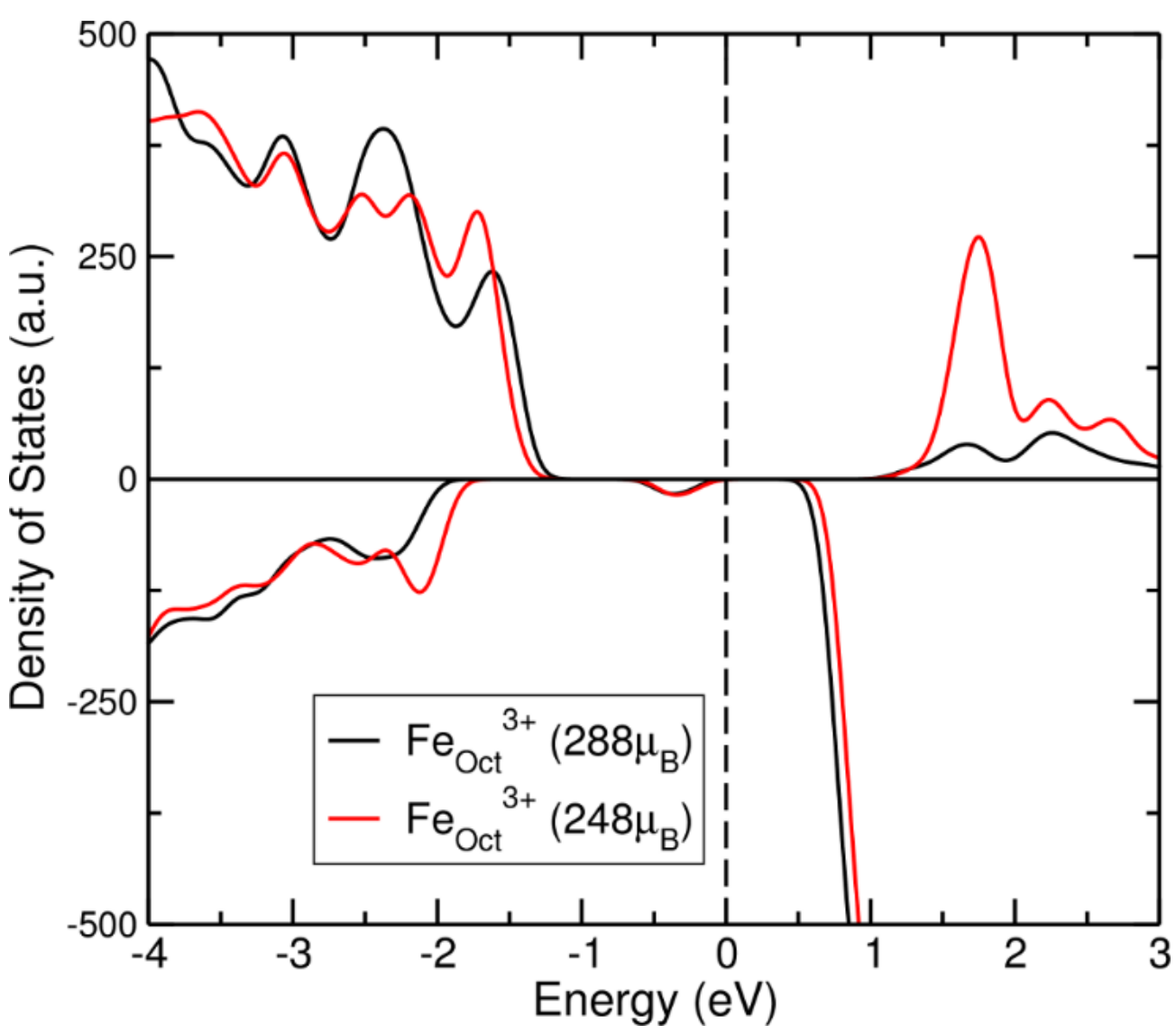

**Figure S2.** PDOS on $Fe_{Oct}^{3+}$ for the naked NP. The black and the red curves represent the projection for the non-spin-flipped system and for the system with all (four) equivalent $Fe_{Oct}^{3+}6c-(\mathbf{3})$ spin-flipped, respectively.

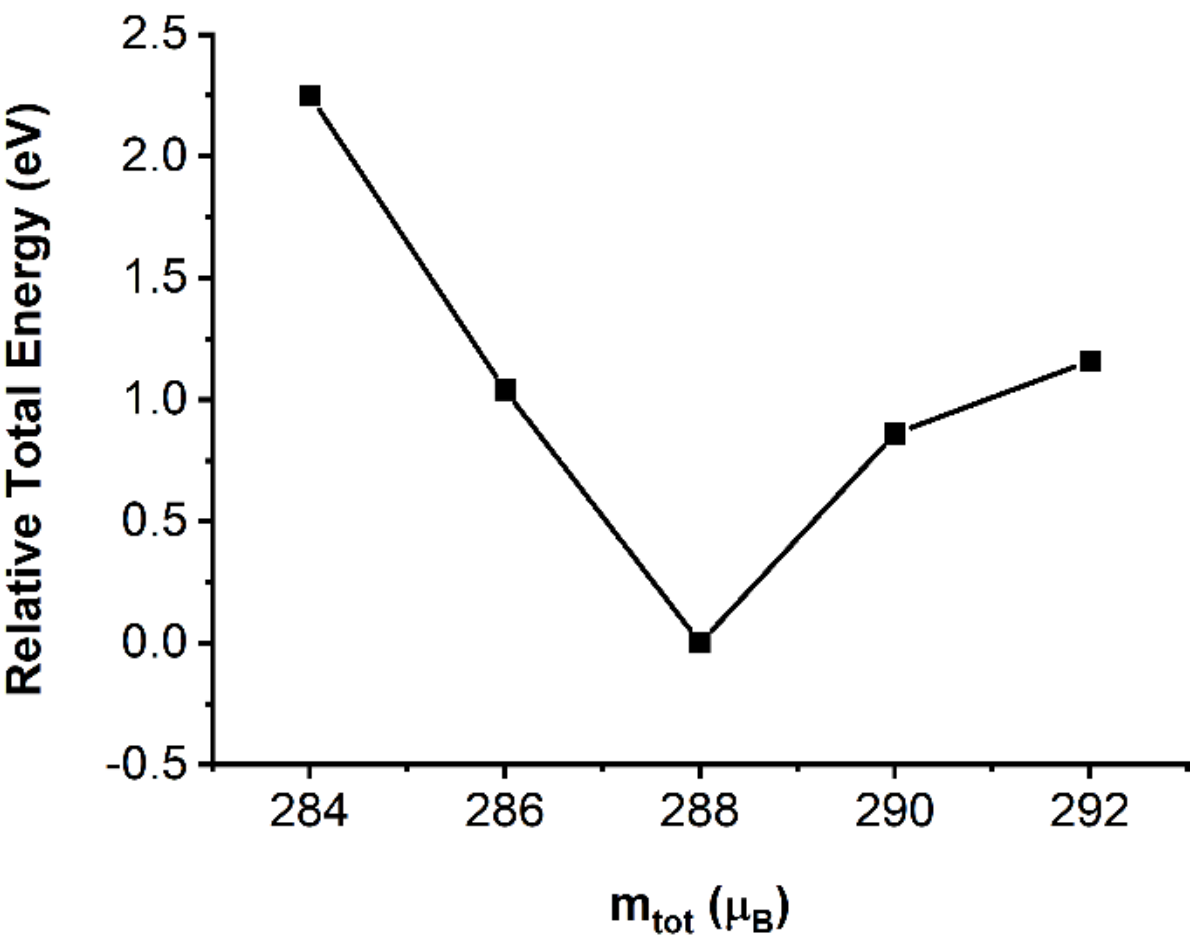

**Figure S3.** The relative total energy as a function of the total magnetic moment ($m_{tot}$) for the $Fe_3O_4$ NP in vacuum. For each $m_{tot}$ value the system is fully relaxed with HSE06 method.

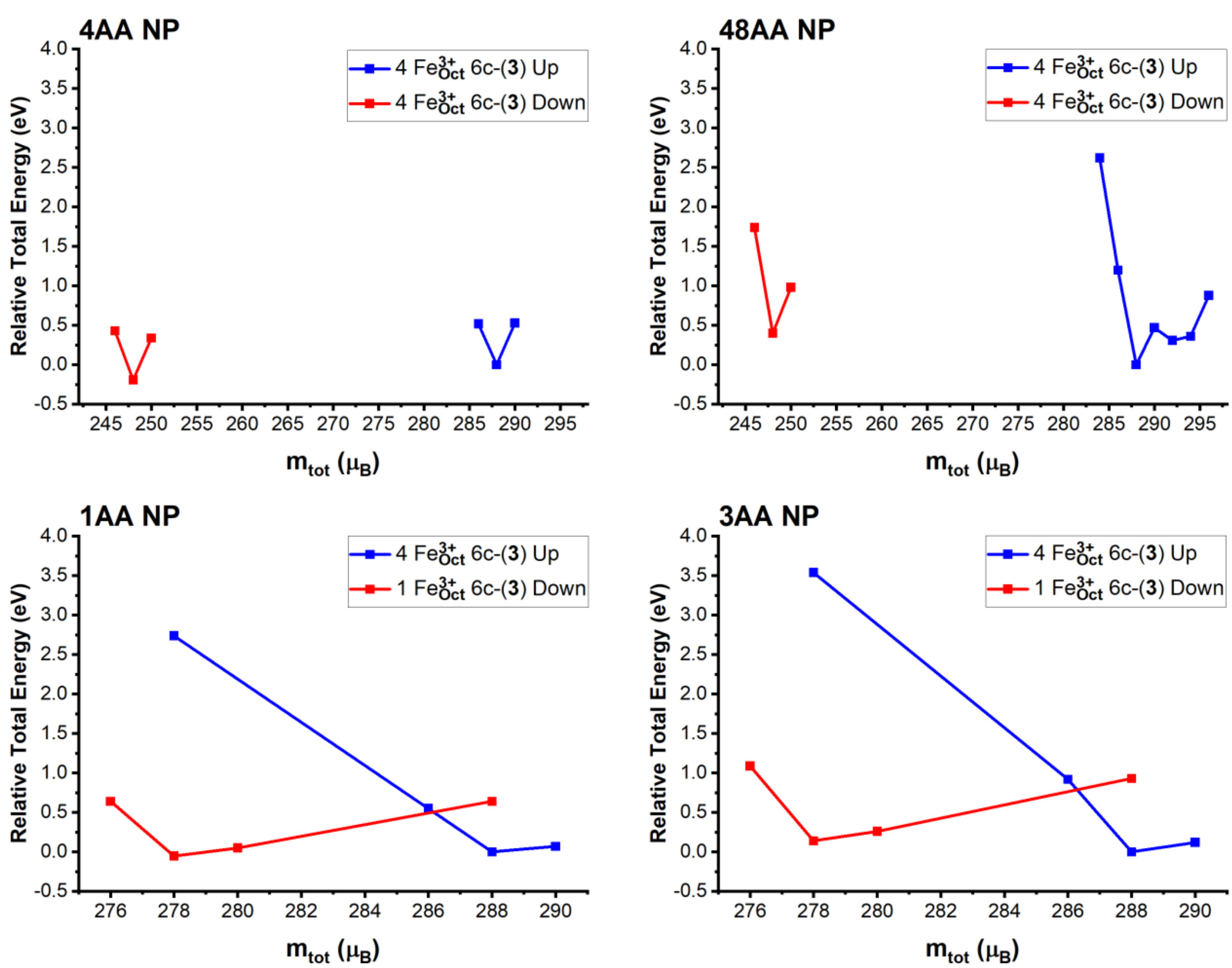

**Figure S4.** The relative total energy as a function of the total magnetic moment ($m_{tot}$) for the $Fe_3O_4$ NP at different acetic acid (AA) coverage: low coverage at corners (4AA NP), full coverage (48AA NP), low coverage (1AA NP), and high coverage at one corner (3AA NP). The blue and the red curves represent the non-spin-flipped and the spin-flipped system, respectively. For each $m_{tot}$ value the system is fully relaxed with HSE06 method.

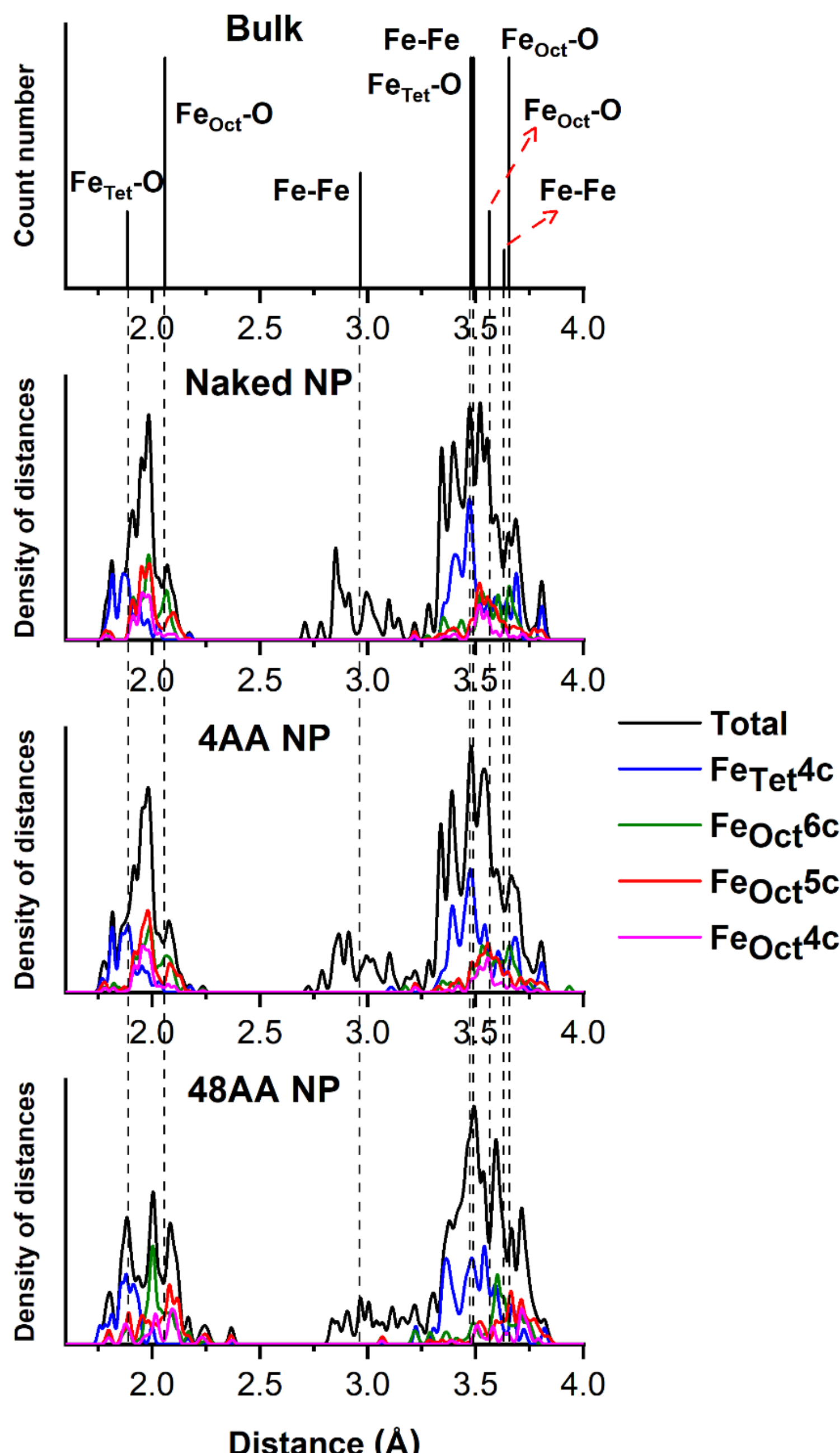

**Figure S5.** Simulated extended X-ray adsorption fine structure (EXAFS) spectra for $Fe_3O_4$ bulk (top panel) and $Fe_3O_4$ NP at different acetic acid (AA) coverage (bottom panels): in vacuum (Naked NP), at low coverage at corners (4AA NP), and at full coverage (48AA NP).

The real space EXAFS was simulated by calculating the density of distances for each Fe ions with other (Fe or O) ions and projecting them on octahedral and tetrahedral Fe ions with O. In general, the range of $Fe_{Tet}$-O and $Fe_{Oct}$-O bond lengths is broadened in the case of NPs with respect to bulk magnetite because of the structural distortions and low coordination near the surface. The broader the peaks, the worse the crystallinity of the nanoparticles. Peaks in simulated EXAFS spectra are similar at different AA coverage.

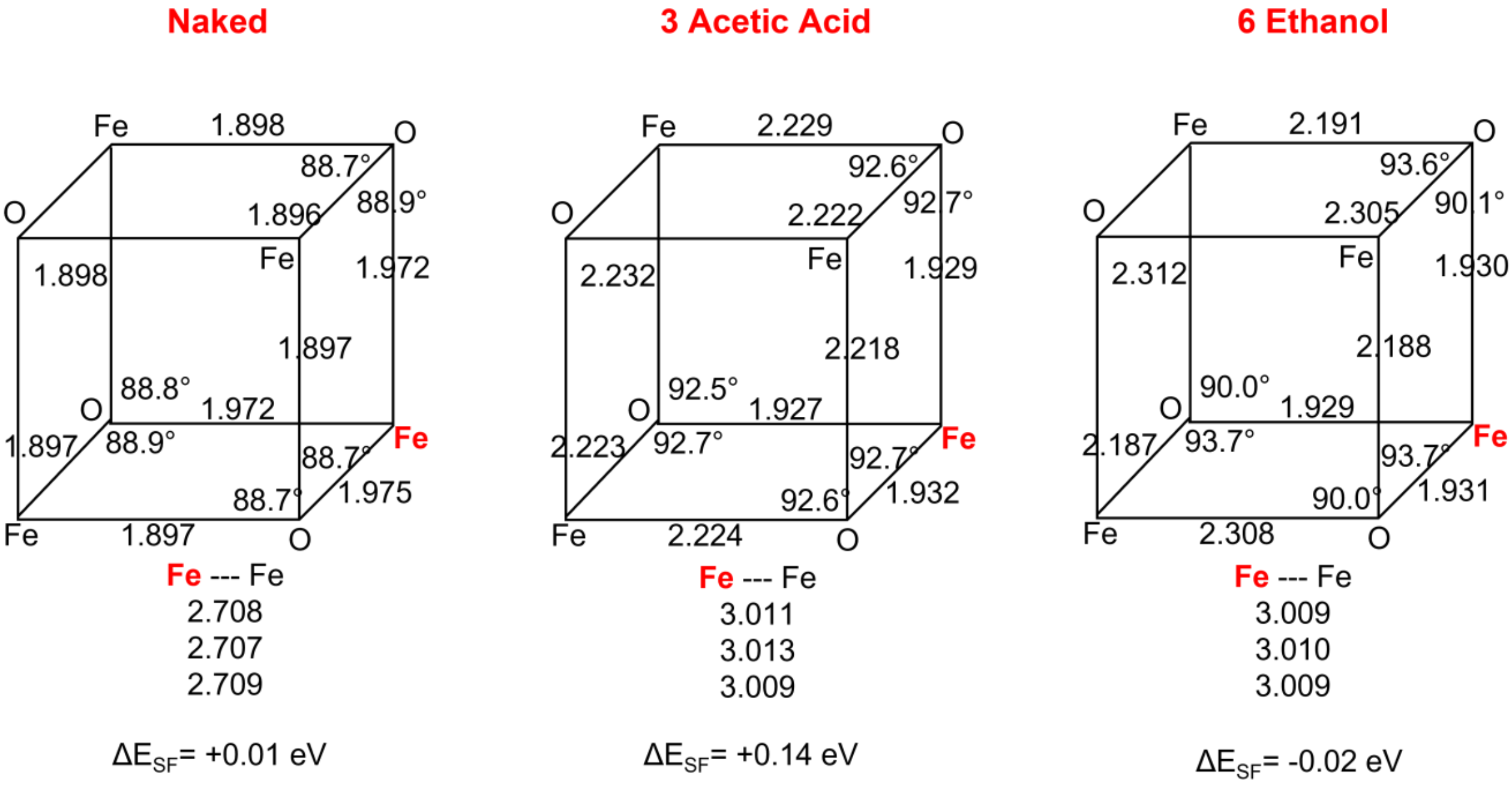

**Figure S6.** Structural (distances and angles) and energetical parameter for the corner of the NP: naked and in presence of additional acetic acid and ethanol molecules. The Fe atom in red is the one called $Fe^{3+}_{Oct}6c-(\mathbf{3})$ in the manuscript, while the black Fe atoms are superficial $Fe_{Oct}$ that interact with the capping molecules. Distances are reported in Angstrom.

The M – O – M angles are in all cases close to 90°. In particular, for the system capped with acetic acid the angles are slightly larger than for the naked, which is an opposite trend with respect to the FM-AFM energy difference, because larger angles are expected to stabilize the AFM configuration, not the FM one [1,2,3]. Regarding the M – O and M---M distances, they are longer in the NP capped with acetic acid than in the naked, again in opposite trend with what expected for a more efficient hopping favouring ferromagnetic superexchange. Moreover, we wish to note that the NP capped with ethanol molecules presents structural parameters very similar to those for the NP capped with acetic acid molecules, but the FM-AFM energy difference is of opposite sign. Therefore, as we stated above, no significant correlations are observed between the variation of the structural parameters and the FM-AFM energy difference in the absence/presence of the acetic acid and ethanol molecules.

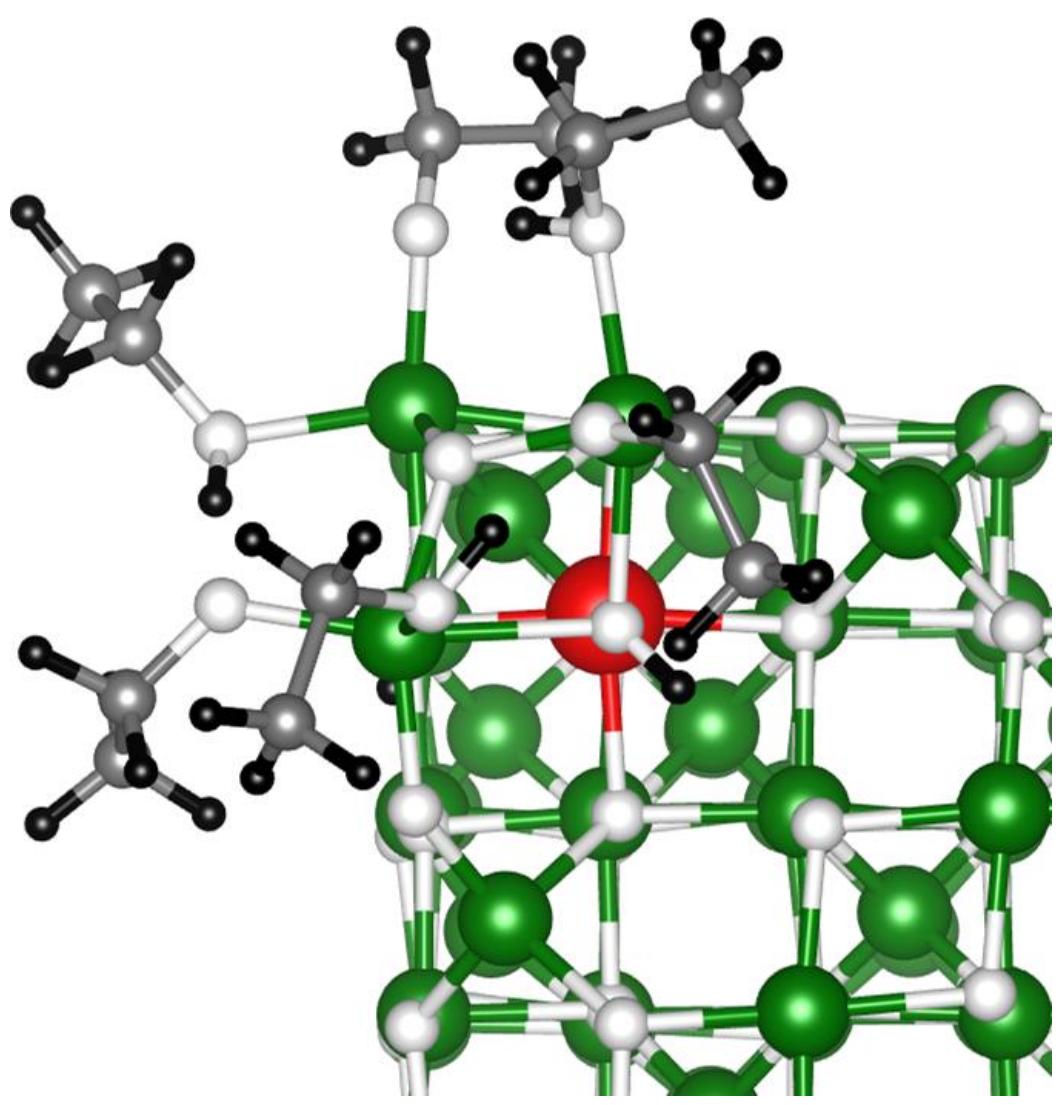

**Figure S7.** Ball and stick representation of the $Fe_3O_4$ cubic NP at ethanol high coverage at one corner (6 ethanol molecules). The black, grey, white, and green beads represent H, C, O, and Fe, respectively. The red bead represents the $Fe_{Oct}$ on which the spin-flip cost is evaluated.

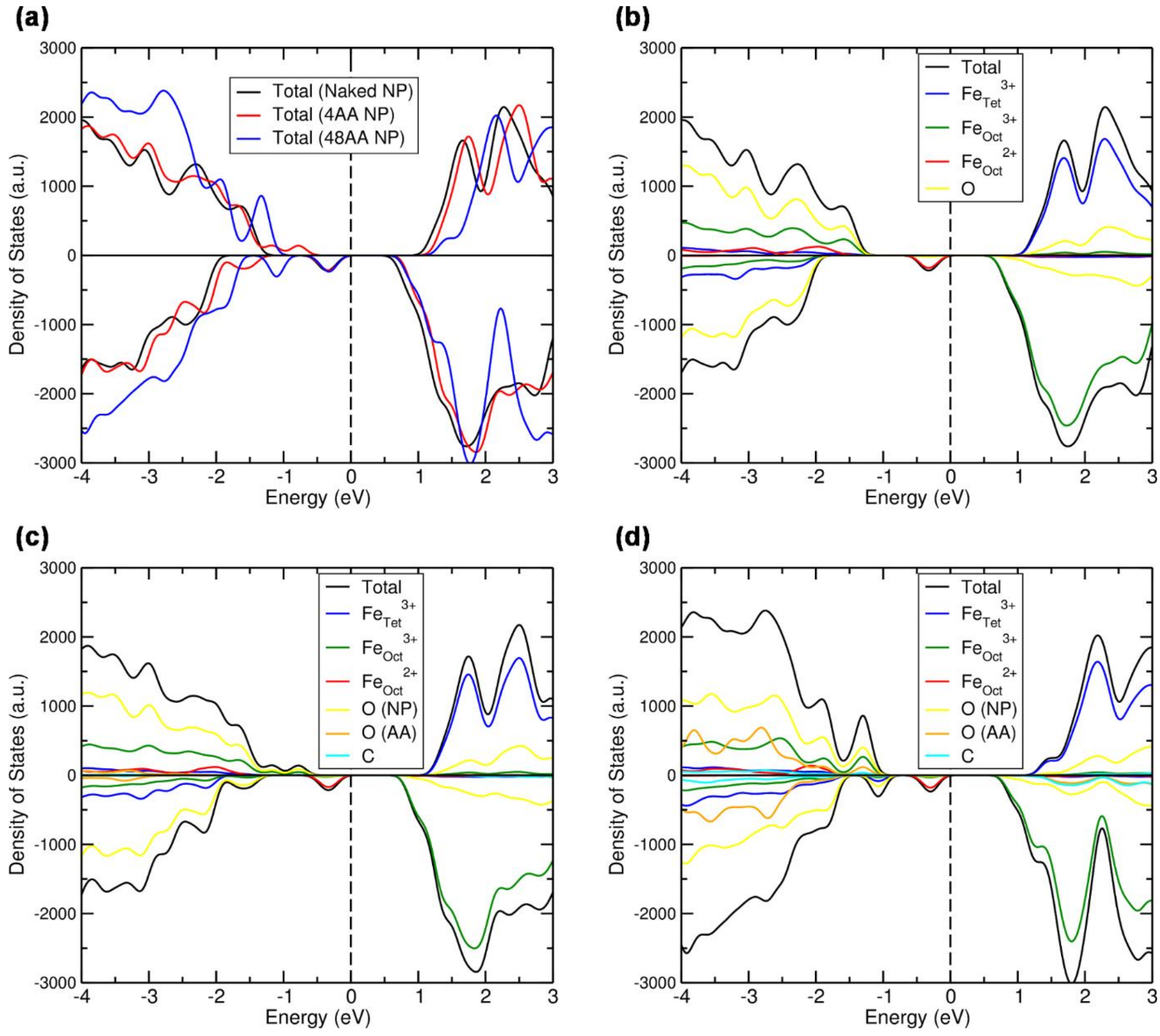

**Figure S8.** (a) TDOS of the NP at different acetic acid (AA) coverage: in vacuum (Naked NP), at low coverage at corners (4AA NP), and at full coverage (48AA NP). PDOS of the (b) naked NP, (c) NP at low coverage at corners (4AA), and (d) NP at full coverage (48AA). The Fermi level is scaled to zero as indicated by the dashed black lines.